\documentclass[amsmath,amssymb,11pt,superscriptaddress,reprint, preprintnumbers, notitlepage,aps,prl,twocolumn]{revtex4-1}
\pdfoutput=1 
\usepackage[utf8]{inputenc}
\usepackage[english]{babel}
\usepackage{amsmath}
\usepackage{graphicx}
\usepackage{pbox}
\usepackage{amssymb}
\usepackage{cancel}
\usepackage{epsfig}
\usepackage[dvipsnames]{xcolor}
\usepackage[normalem]{ulem}
\usepackage{slashed}
\usepackage{soul}
\usepackage{siunitx}
\usepackage{amssymb}
\usepackage{verbatim} %for comment environment
\usepackage{ mathrsfs }
\usepackage{color}
\usepackage[font=small]{caption}
\usepackage[font=small]{subcaption}
\usepackage{url}
\definecolor{MyDarkBlue}{rgb}{0.1, 0.1, 0.8}
\definecolor{SBlue}{rgb}{0.2, 0.4, 0.7} 
\definecolor{MyLightBlue}{rgb}{0.22,0.51,0.9}
\definecolor{MyGreen}{rgb}{0.0, 0.5, 0.0}
\definecolor{BrickRed}{rgb}{0.8, 0.25, 0.33}
\RequirePackage{hyperref}
\hypersetup{colorlinks, citecolor=SBlue,linkcolor=MyDarkBlue, urlcolor=PineGreen}

\newcommand{\eV}{\,\mathrm{eV}}

\newcommand{\MeV}{\,\mathrm{MeV}}
\newcommand{\GeV}{\,\mathrm{GeV}}
\newcommand{\TeV}{\,\mathrm{TeV}}

\makeatletter

\makeatletter
\renewcommand\@makecaption[2]{%
  \par
  \vskip\abovecaptionskip
  \begingroup
  
   \small\rmfamily
    \begingroup
     \samepage
     \flushing
     \let\footnote\@footnotemark@gobble
     \@make@capt@title{#1}{#2}\par
    \endgroup
  \endgroup
  \vskip\belowcaptionskip
}
\makeatother

\DeclareUnicodeCharacter{2212}{-}
\begin{document}

\preprint{MS-TP-24-12}
%%%%%%%%%%%%%%%%%%%%%%%%%%%%%%
\title{\vspace{1cm}\large 
Neutrino masses and mixing from \textit{milli-charged} dark matter}
%%%%%%%%%%%%%%%%%%%%%%%%%%%%%%       
\author{\bf Sudip Jana}
\email[E-mail:]{sudip.jana@mpi-hd.mpg.de}
\affiliation{Max-Planck-Institut f{\"u}r Kernphysik, Saupfercheckweg 1, 69117 Heidelberg, Germany}
%%%%%%%%%%%%%%%%%%%%%%%%%%%%%%
\author{\bf Michael Klasen}
\email[E-mail:]{michael.klasen@uni-muenster.de}
\affiliation{Institut für Theoretische Physik, Universität Münster, Wilhelm-Klemm-Straße 9, 48149 Münster, Germany}
%%%%%%%%%%%%%%%%%%%%%%%%%%%%%%

%%%%%%%%%%%%%%%%%%%%%%%%%%%%%%
\author{\bf Vishnu P.K.}
\email[E-mail:]{ vishnu.pk@uni-muenster.de}
\affiliation{Institut für Theoretische Physik, Universität Münster, Wilhelm-Klemm-Straße 9, 48149 Münster, Germany}
%%%%%%%%%%%%%%%%%%%%%%%%%%%%%%

%%%%%%%%%%%%%%%%%%%%%%%%%%%%%%
\author{\bf Luca Paolo Wiggering}
\email[E-mail:]{luca.wiggering@uni-muenster.de}
\affiliation{Institut für Theoretische Physik, Universität Münster, Wilhelm-Klemm-Straße 9, 48149 Münster, Germany}
%%%%%%%%%%%%%%%%%%%%%%%%%%%%%%

\begin{abstract}
We propose a simple extension to the Standard Model, wherein neutrinos naturally attain small Majorana masses through a one-loop radiative mechanism featuring particles within the loops characterized by milli-charges. Unlike the conventional scotogenic model, our approach avoids imposing a discrete symmetry or expanding the gauge sector. The minuscule electric charges ensure the stability of the lightest particle within the loop as a viable dark matter candidate. Our investigation systematically scrutinizes the far-reaching phenomenological implications arising from these minuscule charges.
\end{abstract}
\maketitle
\section{Introduction}
The electric charges of all known elementary particles are integer multiples of $Q_e/3$, where $Q_e$ represents the electron charge. The apparent quantization of electric charge remains a persistent puzzle in particle physics \cite{Marinelli:1983nd, Dylla:1973zz, Gahler:1982xi, Kyuldjiev:1984kz, Davidson:2000hf}. Developing a theoretical understanding of this quantization is crucial for comprehending the interactions of elementary particles. However, it is theoretically possible for particles to have electric charges of $\epsilon Q_e$, where $\epsilon$ is any real number less than 1. These are referred to as \textit{milli-charged} particles.
Milli-charged particles can be incorporated into the Standard Model through several mechanisms. One well-studied approach introduces a new dark gauge symmetry $U(1)'$, where mixing between the photon and the dark photon allows particles charged under $U(1)'$ to have a small coupling to the photon, resulting in a milli-charge \cite{Holdom:1985ag}. Another method allows Standard Model neutrinos to have tiny charges consistent with non-standard electric charges, given by $Q_{st} + \epsilon (L_i - L_j)$, where $i, j = e, \mu, \tau \ (i \neq j)$, $Q_{st}$ is the standard electric charge, $L_i$ is a family-lepton number, and $\epsilon$ is an arbitrary parameter.
These non-standard charges are compatible with the Standard Model Lagrangian and preserve the gauge anomaly cancellation of $SU(3)_c \otimes SU(2)_L \otimes U(1)_Y$ \cite{Foot:1990uf, Babu:1992sw}. 

In contrast, without introducing any additional symmetries and preserving the charges of Standard Model particles, we extend the SM by adding vector-like fermions or scalars with a small fractional charge $\epsilon$. The anomaly cancellation of the SM is maintained because the fermions are vector-like.
These vector-like fermions and scalars will manifest as \textit{milli-charged} particles within the theory. Our focus is on determining whether such \textit{milli-charged} particles can address some of the significant unresolved issues in particle physics. Specifically, we are interested in exploring whether these particles could serve as candidates for dark matter (DM), given the compelling evidence supporting the existence of DM as a strong indication for physics beyond the Standard Model~\cite{Bertone:2004pz,Klasen:2015uma}. Additionally, considering the well-established phenomenon of neutrino flavor oscillations confirmed by numerous experiments~\cite{Super-Kamiokande:1998kpq,SNO:2001kpb,SNO:2002tuh}, we investigate whether \textit{milli-charged} DM could play a crucial role in neutrino mass generation. 

Our aim is to address the two major challenges of neutrino mass generation and DM stability within a unified framework employing \textit{milli-charged} particles. The \textit{scotogenic model} provides an elegant solution by introducing a symmetry that enables both the suppression of neutrino masses through loop processes and the stabilization of DM \cite{Tao:1996vb, Ma:2006km}. Traditionally, this approach necessitates additional discrete symmetries \cite{Tao:1996vb, Krauss:2002px, Ma:2006km, Aoki:2008av, Ma:2008cu, Restrepo:2013aga, Gustafsson:2012vj, Fiaschi:2018rky, Esch:2018ccs, Jana:2020joi}, which are vulnerable to gravitational anomalies, or requires the extension of the SM gauge group  \cite{Ma:2013yga, Ho:2016aye, deBoer:2023phz, Jana:2019mez, Jana:2019mgj}.
We propose a minimal extension to the Standard Model in which neutrinos naturally acquire small Majorana masses via a one-loop radiative seesaw mechanism involving \textit{milli-charged} particles. Unlike the conventional scotogenic model, our method does not require discrete symmetries or the expansion of the gauge sector. The tiny electric charges ensure the stability of the lightest particle in the loop, making it a viable candidate for DM.

In our model, we propose that the lightest \textit{milli-charged} particle, a vector-like fermion $F$, can serve as a viable DM candidate. Its minuscule electric charge ensures its stability and allows for DM-photon coupling at tree level. However, a purely photon-portal \textit{milli-charged} DM model is inconsistent with current observations \cite{McDermott:2010pa}. The Planck satellite's CMB data rules it out because \textit{milli-charged} DM particles would couple with the baryon-photon plasma even at low temperatures, affecting CMB acoustic peaks and damping DM density fluctuations. Considering the freeze-in mechanism, where $F$ is initially out of equilibrium with the SM thermal bath but produced later through  feeble interactions, generally works for sub-GeV masses \cite{Dvorkin:2019zdi}. However, if \textit{milli-charged} DM also contributes to neutrino mass generation, significant Yukawa interactions are needed, which immediately establish thermal equilibrium between dark matter and the SM bath given a negligible dark matter abundance after reheating, making the freeze-in mechanism incompatible with our model. Therefore, while \textit{milli-charged} DM can generate neutrino masses and mixing successfully, it must annihilate into SM particles via gauge or leptonic portals through the freeze-out scenario to accurately reproduce the relic density. We analyze this scenario for \textit{milli-charged} DM from sub-GeV to TeV mass regimes, discussing various phenomenological consequences such as lepton flavor-violating observables and electroweak precision constraints. Additionally, the presence of \textit{milli-charged} DM modifies the amplitude of $0\nu \beta \beta$ decay, with implications for current and future experiments.

The structure of the paper is as follows: First, in Sec.~\ref{sec:model}, we present the model. Subsequently, in Sec.~\ref{sec:constraints}, we analyze the \textit{milli-charged} DM scenarios for both freeze-in and freeze-out mechanisms, discussing the current constraints and various phenomenological implications. Following this, we introduce in Sec.~\ref{sec:DM} the DM scenario specific to our model, analyze its phenomenology demonstrating its capability to generate neutrino masses and mixing successfully. The resulting implications on the neutrinoless double beta decay are presented in Sec.~\ref{sec:nubeta}. We conclude in Sec.~\ref{sec:conclusion}.
%%%%%%%%%%%%%%%%%%%%%%%%%%%%%%%%%%%%%%%%%%%%%%%%%%%%%%
%%%%%%%%%%%%%%%%%%%%%%%%%%%%%%%%%%%%%%%%%%%%%%%%%%%%%%

\section{The Model \label{sec:model}}
We extend the SM with a single  generation of a \textit{milli-charged} fermion $F$ and two electroweak (EW) scalar doublets $\phi_{1,2}$. No additional symmetries are imposed. The quantum numbers of these multiplets under the SM gauge symmetry $SU(3)_c \otimes SU(2)_L \otimes U(1)_Y$ are given below
\begin{align}
&F_{R,L}\sim(1,1,\epsilon),  \\
&\phi_1 = \begin{pmatrix} \phi_1^{1+\epsilon}  \\ \phi_1^{\epsilon}  \end{pmatrix}\sim (1,2,\dfrac{1}{2}+\epsilon),  \\ 
&\phi_2 = \begin{pmatrix} \phi_2^{1-\epsilon}  \\ \phi_2^{-\epsilon}  \end{pmatrix}\sim (1,2,\dfrac{1}{2}-\epsilon)\,.
\end{align}
Among the five BSM states, three of them $\{F, \phi_1^{\epsilon},\phi_2^{\epsilon}\}$ carry electric charge $\epsilon$. The lightest of these \textit{milli-charged} particles can be a potential candidate for DM, where its stability is guaranteed by its minuscule electric charge. 

The Yukawa interactions of $\{F,\phi_1,\phi_2\}$ and bare mass of $F$ are given by 
\begin{align}
-\mathcal{L}_{\rm Yuk} \supset &Y_1 \overline{F}_{R} \widetilde{\phi}_1^{\dagger} \ell_L + Y_2 \overline{\ell^c_L} (i\sigma_2) \phi_2 F_L 
+ m_F \overline{F}_{L}F_R + h.c. 
\label{Yuk1}
\end{align}
Here, $\ell_L\equiv(\nu_L,e_L)^{T}$ is the SM lepton doublet, while the corresponding charge conjugate field is denoted as $\ell^c_L$, $\widetilde{\phi}_1\equiv i\sigma_2\phi^*_1$, and $\sigma_2$ denotes the second Pauli matrix. 

The scalar potential of the model can be written as
\begin{align}
V(H,\phi_1,&\phi_2)=-\mu_{H}^2H^{\dagger}H + \mu_{\phi_1}^2\phi_1^{\dagger}\phi_1 + \mu_{\phi_2}^2\phi_2^{\dagger}\phi_2
\nonumber\\&
+\frac{\lambda_H}{2}(H^{\dagger}H )^2
+\frac{\lambda_{\phi_1}}{2}(\phi_1^{\dagger}\phi_1)^2
+\frac{\lambda_{\phi_2}}{2}(\phi_2^{\dagger}\phi_2)^2 
\nonumber\\&
+\lambda_{H\phi_1}(H^{\dagger}H)(\phi_1^{\dagger}\phi_1)
+\lambda_{H\phi_1}'(H^{\dagger}\phi_1)(\phi_1^{\dagger}H)
\nonumber\\&
+\lambda_{H\phi_2}(H^{\dagger}H)(\phi_2^{\dagger}\phi_2)
+\lambda_{H\phi_2}'(H^{\dagger}\phi_2)(\phi_2^{\dagger}H)
\nonumber\\&
+\left\{\lambda_{H\phi_1 \phi_2}(H^{\dagger}\phi_1)(H^{\dagger}\phi_2) + h.c.  \right\}
\nonumber\\&
+\lambda_{\phi_1\phi_2}(\phi_1^{\dagger}\phi_1)(\phi_2^{\dagger}\phi_2)
+\lambda_{\phi_1\phi_2}'(\phi_1^{\dagger}\phi_2)(\phi_2^{\dagger}\phi_1). 
\label{eq:pot}
\end{align}
Here, $H\equiv (0, \frac{v+h}{\sqrt{2}})^T$ represents the SM Higgs doublet after EW symmetry braking with $v\simeq 246 \GeV$ denoting the EW vacuum expectation value and $h$ is the SM Higgs. The quartic coupling $\lambda_{H\phi_1 \phi_2}$ leads to mixing between the \textit{milli-charged} 
scalars $\phi_1^{\epsilon}$ and $\phi_2^{\epsilon}$, resulting in the following mass squared matrix:
\begin{align}
\mathcal{M}^2_{\phi^{{\epsilon}}}=
\begin{pmatrix}
\mu^2_{\phi_1}+\frac{(\lambda_{H\phi_1}+\lambda'_{H\phi_1})}{2}v^2  &  \frac{\lambda_{H \phi_1 \phi_2} }{2}v^2\\
   \frac{\lambda_{H \phi_1 \phi_2} }{2}v^2 &  \mu^2_{\phi_2}+\frac{(\lambda_{H\phi_2}+\lambda'_{H\phi_2})}{2}v^2
\end{pmatrix}. \label{S0}
\end{align}
The corresponding mass eigenstates $S_{1,2}^{{\epsilon}}$ with masses $m_{S_{1,2}}$ are related to $\phi_{1,2}^{{\epsilon}}$ through the following mixing matrix
\begin{equation}\label{mix}
    \begin{pmatrix} S_1^{{\epsilon}} \\ S_2^{{\epsilon}}\end{pmatrix}
    =\begin{pmatrix}
    \cos{\theta}  & -\sin{\theta} \\
	\sin{\theta} & \cos{\theta}
    \end{pmatrix}\begin{pmatrix} \phi_1^{{\epsilon}}\\ \phi_2^{{\epsilon}} \end{pmatrix} \,.
\end{equation}
Here, $\theta$ is the mixing angle, which can be related to the quartic coupling $\lambda_{H \phi_1 \phi_2}$ via
\begin{equation} \sin{2\theta}=\dfrac{\lambda_{H \phi_1 \phi_2} v^2 }{m^2_{S_2}-m^2_{S_1}}.
\end{equation}
The masses of the scalars $\phi_1^{1+{\epsilon}}$ and $\phi_2^{1-{\epsilon}}$ are denoted as $m_{\phi_1}$ and $m_{\phi_2}$, respectively, and are given by 
\begin{align}
m^2_{\phi_1}=\mu^2_{\phi_1}+\frac{\lambda_{H\phi_1}}{2}v^2, \quad  m^2_{\phi_2}=\mu^2_{\phi_2}+\frac{\lambda_{H\phi_2}}{2}v^2.
\end{align}
%%%%%%%%%%%%%%%%%%%%%%%%%%%%%%%%%%%%%%%%%%%%%%%%%%%%%%
%%%%%%%%%%%%%%%%%%%%%%%%%%%%%%%%%%%%%%%%%%%%%%%%%%%%%%
\begin{figure}[thb!]
\includegraphics[width=0.32\textwidth]{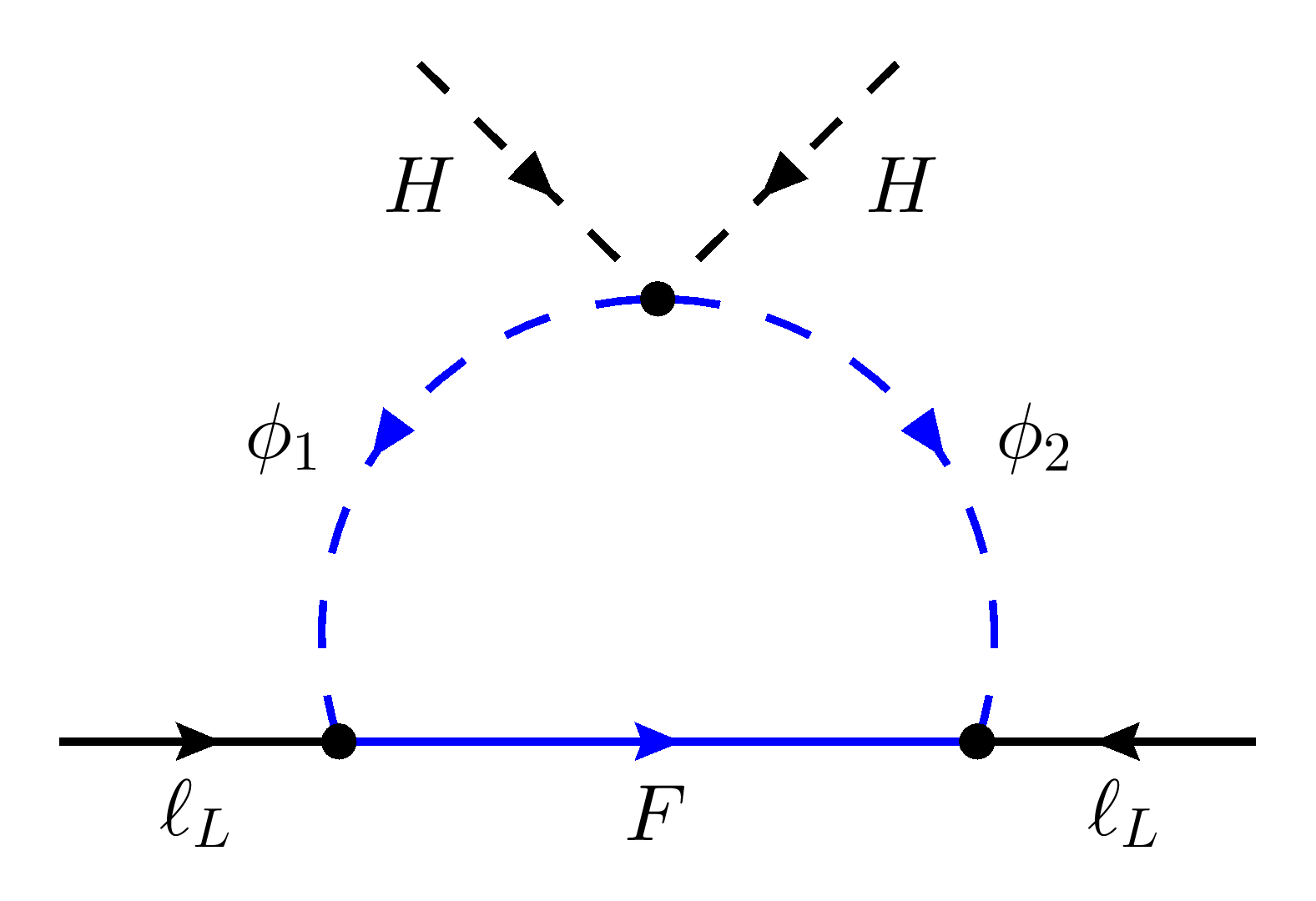}
\caption{Feynman diagram for neutrino masses and mixings generated by \textit{milli-charged} particles. 
}\label{neutmass}
\end{figure}
%%%%%%%%%%%%%%%%%%%%%%%%%%%%%%%%%%%%%%%%%%%%%%%%%%%%%%%%%%%%%%%%%%%%%

The simultaneous presence of the Lagrangian terms in Eq.~\eqref{Yuk1} and the quartic coupling term $\lambda_{H\phi_1 \phi_2}(H^{\dagger}\phi_1)(H^{\dagger}\phi_2) + h.c $ leads to lepton number violation by two units, which results in radiative neutrino mass generation at the one-loop level via the diagram depicted in Fig.~\ref{neutmass}. The corresponding neutrino mass matrix is given by 
\begin{align}
\label{eq:mnu}
\mathcal{M}^\nu&=m_0\left(Y_1.Y_2^T+    Y_2.Y_1^T\right), 
\end{align}
where
\begin{align}
m_0&=\frac{m_{F}\sin 2\theta}{32\pi^2} 
\left\{\frac{m_{S_2}^2\ln\left( \frac{m_{S_2}^2}{m^2_F} \right) }{m_{S_2}^2-m_F^2}
-
\frac{m_{S_1}^2\ln\left( \frac{m_{S_1}^2}{m^2_F} \right) }{m_{S_1}^2-m_F^2}
\right\}.
\end{align}
The above mass matrix can be diagonalized by a unitary transformation of the form $U^T \mathcal{M}^{\nu} U$, where $U$ denotes the PMNS matrix. Note that $\mathcal{M}^{\nu}$ is of rank 2, which implies that the lightest neutrino is massless in our model. Moreover, the particular form of $\mathcal{M}^{\nu}$ allows one to parameterize the Yukawa couplings $Y_{1,2}$ in terms of the elements of the PMNS matrix, two non-zero neutrino masses $m_i$, and masses and mixings of the BSM states that enter in the loop diagram shown in Fig.~\ref{neutmass}. The corresponding parametrization for the normal hierarchy (NH) of neutrino masses is given by \cite{Cordero-Carrion:2018xre,Cordero-Carrion:2019qtu,Hagedorn:2018spx}
\begin{align}
&Y_1=\sqrt{\frac{k}{2m_0}}\left(
\begin{array}{c}
 \sqrt{m_3} U_{13}^{*} + i \sqrt{m_2} U_{12}^{*} \\\sqrt{m_3} U_{23}^{*} + i \sqrt{m_2} U_{22}^{*} \\ \sqrt{m_3} U_{33}^{*} + i \sqrt{m_2} U_{32}^{*} \\
\end{array}
\right),\nonumber \\
&Y_2=\sqrt{\frac{1}{2m_0 k}}\left(
\begin{array}{c}
 \sqrt{m_3} U_{13}^{*} - i \sqrt{m_2} U_{12}^{*} \\\sqrt{m_3} U_{23}^{*} - i \sqrt{m_2} U_{22}^{*} \\ \sqrt{m_3} U_{33}^{*} - i \sqrt{m_2} U_{32}^{*} \\
\end{array}
\right),
\label{eq:YukpNH}
\end{align}
where $k$ signifies the relative hierarchy between the Yukawa couplings $Y_1$ and $Y_2$. For the inverted hierarchy (IH) of neutrino masses, the parametrization takes the following form
\begin{align}
&Y_1=\sqrt{\frac{k}{2m_0}}\left(
\begin{array}{c}
 \sqrt{m_1} U_{11}^{*} + i \sqrt{m_2} U_{12}^{*} \\\sqrt{m_1} U_{21}^{*} + i \sqrt{m_2} U_{22}^{*} \\ \sqrt{m_1} U_{31}^{*} + i \sqrt{m_2} U_{32}^{*} \\
\end{array}
\right), \nonumber \\
&Y_2=\sqrt{\frac{1}{2m_0 k}}\left(
\begin{array}{c}
 \sqrt{m_1} U_{11}^{*} - i \sqrt{m_2} U_{12}^{*} \\\sqrt{m_1} U_{21}^{*} - i \sqrt{m_2} U_{22}^{*} \\ \sqrt{m_1} U_{31}^{*} - i \sqrt{m_2} U_{32}^{*} \\
\end{array}
\right).
\label{eq:YukpIH}
\end{align}
For our analysis, we use the above parametrizations for $Y_1$ and $Y_2$.
%%%%%%%%%%%%%%%%%%%%%%%%%%%%%%%%%%%%%%%%%%%%%%%%%%%%%%
%%%%%%%%%%%%%%%%%%%%%%%%%%%%%%%%%%%%%%%%%%%%%%%%%%%%%%
\medskip
\section{Constraints on the millicharged dark matter \label{sec:constraints}}
\begin{figure}[thb!]
\includegraphics[width=0.4\textwidth]{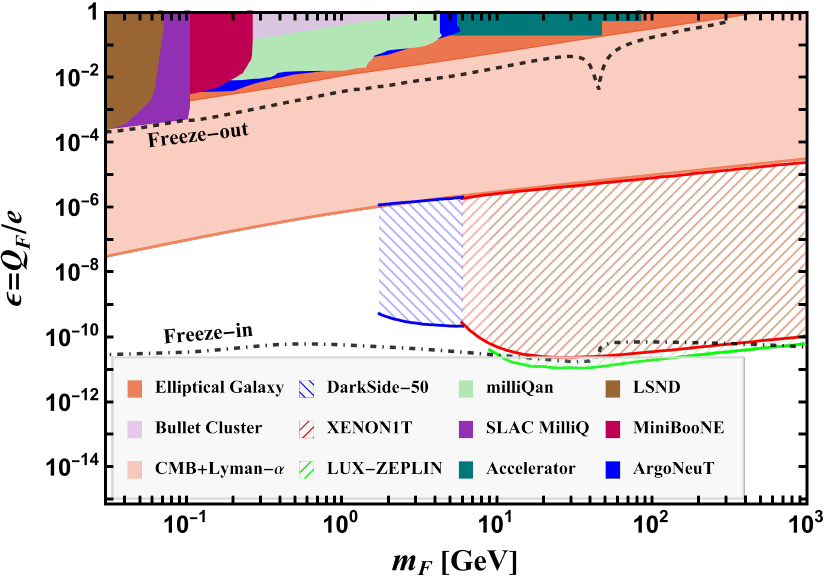}
\caption{ Current status of \textit{milli-charged} DM in $m_F$-$\epsilon$ plane.  The bounds are imposed by the following constraints: CMB+Lyman-$\alpha$ constraints \cite{Xu:2018efh,Planck:2015fie,SDSS:2004kjl} (light-orange region),  DM halo constraints from the bullet cluster \cite{McDermott:2010pa,Ibe:2009mk} (light-purple region) and elliptical galaxies \cite{McDermott:2010pa,Markevitch:2003at} (orange region), MilliQAn demonstrator \cite{Citron:2021woe,Ball:2020dnx} (light-green region), SLACmQ \cite{Prinz:1998ua,Davidson:2000hf} (purple region),  LEP + beam-dump experiments \cite{Davidson:2000hf} (green region), LSND \cite{Magill:2018tbb,LSND:1996jxj} (brown region), MiniBooNE \cite{Magill:2018tbb,MiniBooNE:2018esg} (maroon region), ArgoNeuT \cite{ArgoNeuT:2019ckq} (blue region), and DM direct detection constraints from DarkSide-50 \cite{DarkSide:2018bpj,Belanger:2020gnr} (blue shaded region), Xenon1T \cite{XENON:2018voc,Belanger:2020gnr} (red shaded region), and Lux-Zeplin \cite{LZ:2022lsv} (green shaded region). The dashed and dot-dashed curves represent the DM relic density $\Omega h^2 =0.12$ for the scenarios of freeze-out and freeze-in mechanisms, respectively. 
See text for more details.    
}\label{mDM:constraints}
\end{figure}

%%%%%%%%%%%%%%%%%%%%%%%%%%%%%%%%%%%%%%%%%%%%%%%%%%%%%%
Before investigating the phenomenological implications of our model, we discuss the constraints on the standard \textit{milli-charged} DM scenario, where the DM couples to the SM particles only via photons. These constraints are summarized in Fig.~\ref{mDM:constraints}. 

%%%%%%%%%%%%%%%%%%%%%%%%%%%%%%%%%%%%%%%%%%%%%
%%%%%%%%%%%%%%%%%%%%%%%%%%%%%%%%%%%%%%%%%%%%%

The interactions of \textit{milli-charged} DM with the SM thermal plasma during the recombination epoch are constrained by the precise measurements of CMB observables. 
Limits on such interactions were derived in Ref. \cite{Xu:2018efh} using  Planck satellite data \cite{Planck:2015fie} and the Sloan Digital Sky Survey \cite{SDSS:2004kjl}, which are indicated in Fig.~\ref{mDM:constraints} through the light-orange shaded region.

There are strong constraints on the electric charge of DM from direct detection experiments  since the charged DM can scatter off the nuclei at  tree level via photon exchange.  The bounds from  DarkSide-50 \cite{DarkSide:2018bpj}  and Xenon1T \cite{XENON:2018voc} derived in Ref. \cite{Belanger:2020gnr},
are depicted in Fig.~\ref{mDM:constraints} by the blue-shaded region and the red-shaded region, respectively. Using these, we recast the limits from Lux-Zeplin \cite{LZ:2022lsv}, which are represented in Fig.~\ref{mDM:constraints} by the green-shaded region.

 %%%%%%%%%%%%%%%%%%%%%%%%%%%%%%%%%%%%%%%%%%%%
\begin{figure*}[t!]
\includegraphics[width=0.85\textwidth]{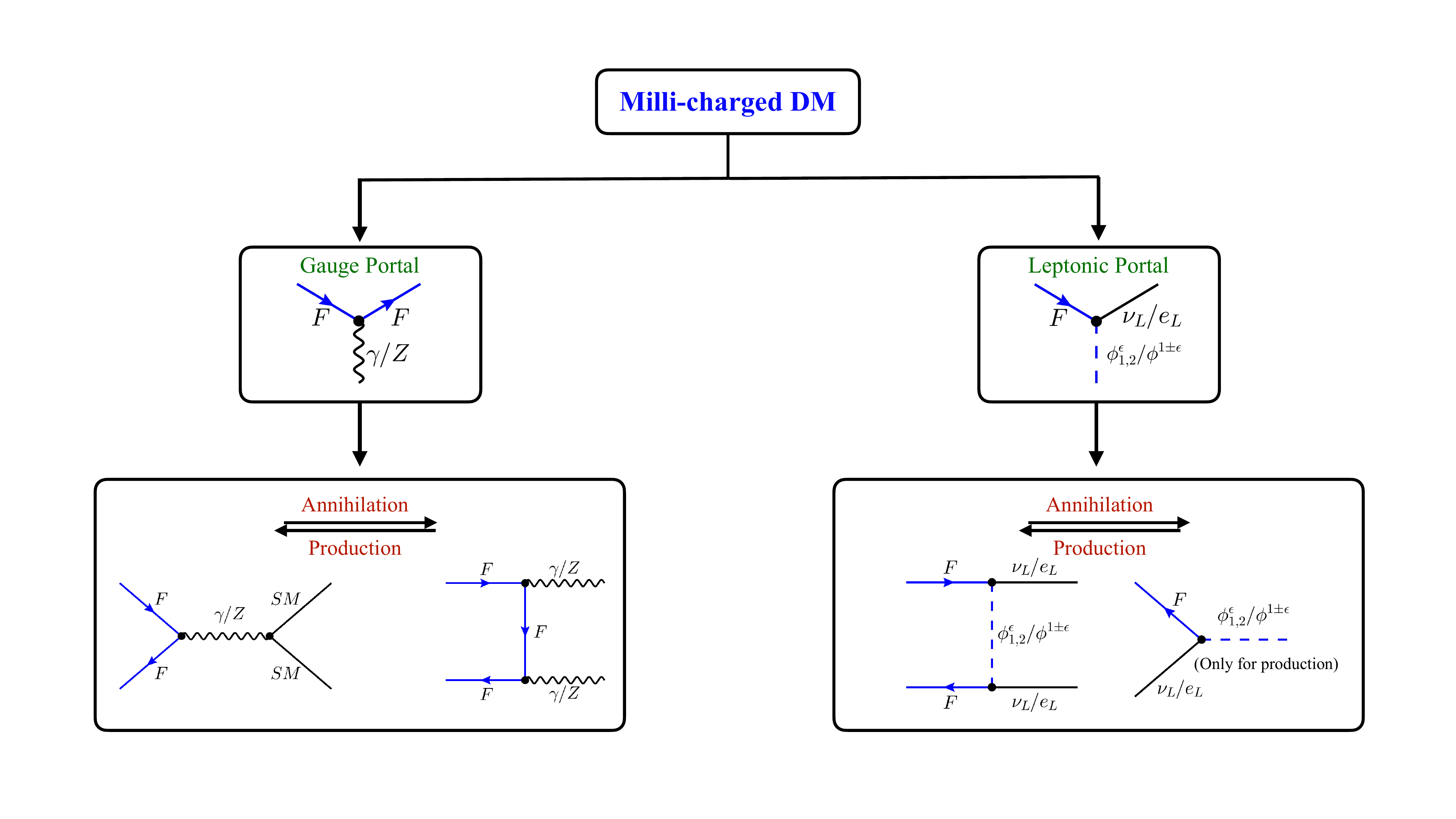}
\caption{Different portals available for DM annihilation/production in our model.  
}\label{fig:Flowchart}
\end{figure*}
%%%%%%%%%%%%%%%%%%%%%%%%%%%%%%%%%%%%%%%%%%%%

At collider or fixed target experiments, \textit{milli-charged} DM can be produced via the Drell-Yan process or via the decay of various mesons. The null results from  the following experiments impose stringent constraints on the  millicharge and are depicted in Fig.~\ref{mDM:constraints}:  MilliQAn demonstrator \cite{Citron:2021woe,Ball:2020dnx} (light-green colored region),  SLACmQ \cite{Prinz:1998ua,Davidson:2000hf} (purple colored region), LEP + beam-dump experiments \cite{Davidson:2000hf} (green colored region),  LSND \cite{Magill:2018tbb,LSND:1996jxj} (brown colored region), MiniBooNE \cite{Magill:2018tbb,MiniBooNE:2018esg} (maroon colored region), and  ArgoNeuT \cite{ArgoNeuT:2019ckq} (blue colored region).

The self-interaction cross section of DM is constrained through observations of dark matter halos like galaxy cluster mergers and the ellipticity of galaxies. The resulting limits from the bullet cluster~\cite{Ibe:2009mk} and the elliptical galaxy NGC 720~\cite{Markevitch:2003at} on the electromagnetic charge of dark matter are derived in Ref. \cite{McDermott:2010pa} and shown in Fig.~\ref{mDM:constraints} in light-purple and orange, respectively.

%%%%%%%%%%%%%%%%%%%%%%%%%%%%%%%%%%%%%%%%%%%%
\section{Dark matter Phenomenology \label{sec:DM}}
%%%%%%%%%%%%%%%%%%%%%%%%%%%%%%%%%%%%%%%%%%%%

In our model, the lightest \textit{milli-charged} particle can be a viable candidate for DM, which we choose to be the vector-like fermion $F$. The minuscule electric charge of $F$ ensures its stability, which also results in tree-level DM-photon coupling. In addition to its gauge interactions, $F$ also has leptonic interactions via the Yukawa couplings given in Eq.~\eqref{Yuk1}. Hence, the \textit{milli-charged} DM candidate in our model is not pure gauge portal DM like in Refs. \cite{Dvorkin:2019zdi, McDermott:2010pa}. 
As discussed earlier, these Yukawa couplings are crucial for explaining neutrino masses and mixings. We will show later in this section that they also play a vital role in DM phenomenology.

We investigate both the freeze-in and the freeze-out mechanisms for explaining the observed amount of DM. The dominant channels that contribute to the relic abundance are shown in Fig.~\ref{fig:Flowchart}.

\subsection{Freeze-in scenario} 
First, we look into the freeze-in scenario. In this case, the DM candidate $F$ is assumed to a have a negligible abundance after reheating but is produced from the SM bath at later times.

In the case of pure gauge coupling-induced processes, 
$F$ is mainly produced via the annihilation of the SM particles (see Fig.~\ref{fig:Flowchart}). The amplitude of these processes depends on the mass of $F$ and its electric charge $\epsilon$. Using  {\tt{micrOMEGAs 5.3.41}} \cite{Belanger:2018ccd}, we calculated the relic abundance for this scenario. The corresponding parameter space consistent with the Planck relic density constraint \cite{Planck:2018vyg} $\Omega h^2 =0.12$ is shown as a dot-dashed line in Fig.~\ref{mDM:constraints}. As one can see,  $\epsilon\sim 10^{-11}$ is required to generate the correct relic abundance. Such values of $\epsilon$ are also consistent with the out-of-equilibrium condition that is required for the successful working of the freeze-in scenario.

However, once the Yukawa couplings are turned on as per Eq. \eqref{eq:YukpNH} or Eq. \eqref{eq:YukpIH}, 
a random parameter scan revealed that 
the DM immediately thermalizes with the SM plasma, rendering 
the neutrino mass mechanism and the freeze-in mechanism  incompatible with each other in our model.

To illustrate this, first, we calculate the limit on the Yukawa couplings imposed by the out-of-equilibrium condition, which read as
\begin{align}
\label{eq:YukBound}
    \Gamma(\phi\xrightarrow{} F \ell) \lesssim H|_{T\simeq m_{\phi}} \Longrightarrow    |Y_{1,2}| \lesssim 10^{-8} \sqrt{\dfrac{m_{\phi_{1,2}}}{100 \, \mathrm{GeV}}}.
\end{align}
Here $H \simeq \frac{T^2}{M_{\mathrm{Pl}}}$ stands for the Hubble expansion rate,  $M_{\mathrm{Pl}}=2.4\times 10^{18}\,\GeV$ is  the reduced Planck mass and $\Gamma$ denotes the production rate of $F$ via the decay of BSM scalars. With this limit on the Yukawa couplings, we estimate the neutrino masses for a scenario where $m_F \ll m_{S_1}\sim m_{\phi_1}, m_{S_2}\sim m_{\phi_2}$ giving
\begin{align}
\label{eq:NeutFIN1}
    M_{\nu} &  \simeq   \frac{Y_1 Y_2 \sin{2\theta}}{16 \pi^2}m_{F} \ln \left(\frac{m_{S_2}^2}{m_{S_!}^2}\right),\,\notag \\
    &\simeq
    \begin{cases}
        10^{-9} \eV \left(\frac{m_F}{1 \GeV} \right)\left(\frac{\sin 2\theta}{1}\right)  & m_{S_{1,2}} \sim \mathcal{O} (100) \GeV,  \\
        10^{-11} \eV \left(\frac{m_F}{1 \GeV} \right)\left(\frac{\lambda_{H \phi_1 \phi_2}}{2}\right) & m_{S_{1,2}} \sim \mathcal{O} (100) \TeV,
    \end{cases}
\end{align}
As one can see,  the resulting neutrino masses are about eight to ten orders of magnitude smaller than the observed values.

\subsection{Freeze-out scenario}
Next, we turn our attention to the freeze-out mechanism. In this case, the relic abundance of DM is mainly governed by its (co)annihilation cross-section into SM final states ($\sigma v$). As aforementioned, in our model, $F$ annihilates into SM particles either via the gauge or the leptonic portal.

The contributions from the gauge portal processes depend on $M_F$ and $\epsilon$.
In Fig.~\ref{mDM:constraints}, we show $\epsilon$ as a function of $M_F$ for $\Omega h^2 =0.12$ as a dashed line. 
As one can see, this parameter space is excluded by the CMB constraints \cite{Xu:2018efh}. 
Moreover, to be consistent with the various constraints on the  electric charge of DM, $\epsilon$ is required to be very small, $\epsilon \ll 1$. Hence, the contribution from gauge portal processes is negligible in our scenario such that we set $\epsilon=10^{-12}$ in the remainder of the paper. 

On the other hand, the contributions that arise from the leptonic portal can be significant. The $s$-wave components of the corresponding cross sections  are given by \cite{Herms:2023cyy}
\begin{align}
\label{eqs:annihilation}
    &(\sigma v)_{S_1}^{\bar F F \to \bar\nu\nu} = \frac{m_F^2}{32\pi}\left(\frac{ (|Y_1|^2c_{\theta}^2+|Y_2|^2 s_{\theta}^2)^2}{(m_F^2 + m_{S_1}^2)^2}\right),\\
    &(\sigma v)_{S_2}^{\bar F F \to \bar\nu\nu} = \frac{m_F^2}{32\pi}\left(\frac{ (|Y_1|^2s_{\theta}^2+|Y_2|^2 c_{\theta}^2)^2}{(m_F^2 + m_{S_2}^2)^2}\right), \\
    &(\sigma v)_{\phi_1}^{\bar F F \to \ell^+ \ell^-} = \frac{m_F^2}{32\pi}\left(\frac{ |Y_1|^4}{(m_F^2 + m_{\phi_1}^2)^2}\right),\\
    &(\sigma v)_{\phi_2}^{\bar F F \to \ell^+ \ell^-} = \frac{m_F^2}{32\pi}\left(\frac{|Y_2|^4}{(m_F^2 + m_{\phi_2}^2)^2}\right),
\end{align}
where the subscripts denote the scalar mediators appearing in the underlying $t$-channel processes.
It is worth mentioning that the annihilation processes mediated by the \textit{milli-charged} scalars $\{S_1^{{\epsilon}},S_2^{{\epsilon}}\}$ result in neutrinos as annihilation products, whereas the processes mediated by $\{\phi_1^{1+{\epsilon}},\phi_2^{1-{\epsilon}}\}$ result in charged leptons.

In the following, we explore separately the low-mass and the high-mass regimes of DM, which we define through the conditions $M_F<1\,\mathrm{GeV}$ and $10\,\mathrm{GeV} < M_F$, respectively.

\textbf{\emph{Light dark matter}.--}
In this case, in order to generate an adequately large contribution to $(\sigma v)$, at least one of the mediator states of annihilation processes needs to be light \cite{Okawa:2020jea, Herms:2023cyy}. Since the masses of the scalars $\{\phi_1^{1+{\epsilon}},\phi_2^{1-{\epsilon}}\}$ are constrained to be above $110$ GeV to satisfy collider bounds \cite{Babu:2019mfe,Iguro:2022tmr, Jana:2020pxx}, 
one of the \textit{milli-charged} scalars $\{S_1^{{\epsilon}},S_2^{{\epsilon}}\}$ has to be light to fulfill this condition. Moreover, the sum $m_{S_1}+m_{S_2}$ needs to be larger than $m_Z$ to avoid $Z$-boson decays into $S_1^{\pm \epsilon}  S_2^{\mp \epsilon}$ which are severely constrained by measurements of its decay width \cite{Electroweak:2003ram,Tanabashi:2018oca}.
Without loss of generality,  $S_1^{{\epsilon}}$ is therefore chosen to be the light mediator state while the mass of $S_2^{{\epsilon}}$ is selected such that $m_{S_1}+m_{S_2}>m_Z$. 
The requirement of $m_{S_1}<1\,\mathrm{GeV}$ and $m_{S_2},m_{\phi_1},m_{\phi_2}\sim \mathcal{O}(100)\, \GeV$ imposes additional constraints on the model parameter space, which we address next.

%%%%%%%%%%%%%%%%%%%%%%%%%%%%%%
\begin{figure*}[t!]
$$
\includegraphics[width=0.4\textwidth]{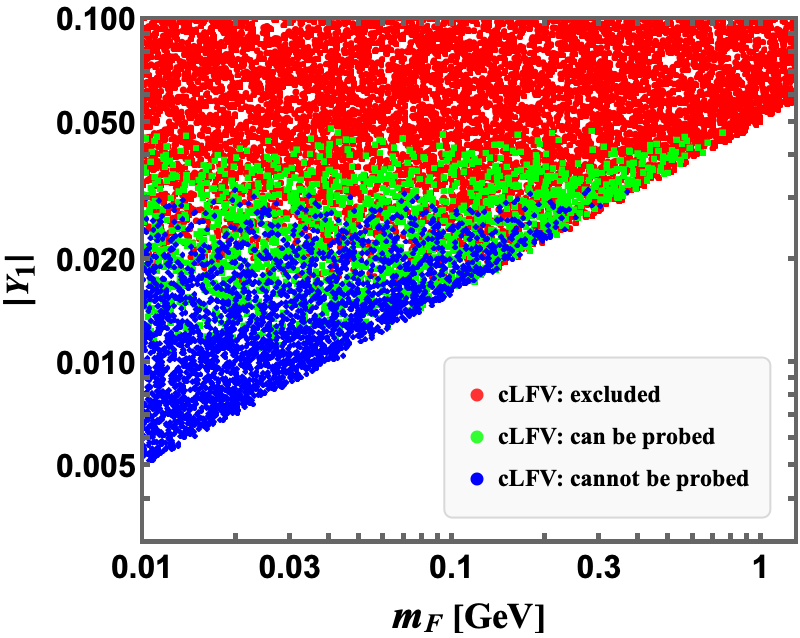}\hspace{0.45in}
\includegraphics[width=0.4\textwidth]{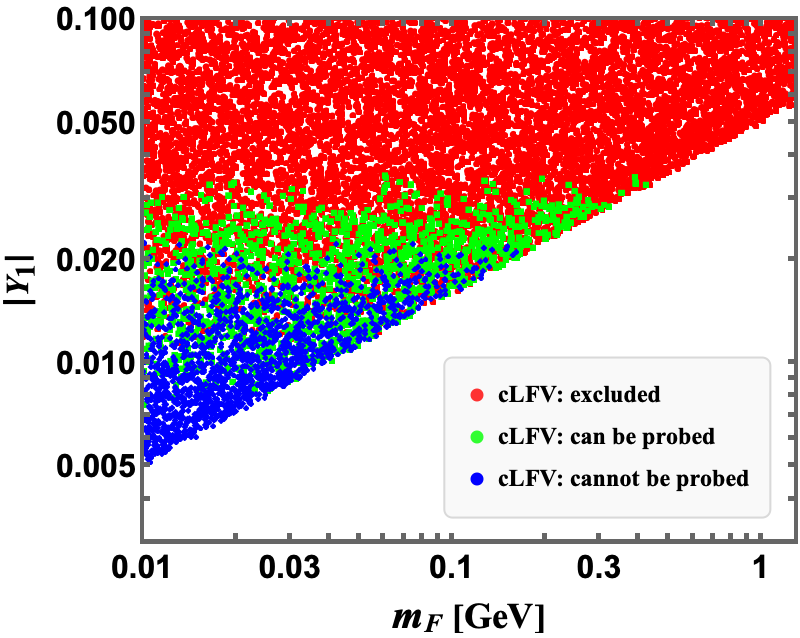}
$$
\caption{Parameter space of the light-DM scenario consistent with the Planck relic density constraint in terms of $(m_F, |Y_1|)$ for both NH (left panel) and IH (right panel) of neutrino masses.  
The red-colored data points are excluded by cLFV constraints given in Eq.~\eqref{eq:cLFV}. The data points represented by green (blue) color can (not) be probed in future cLFV experiments.
}\label{fig:lightDM}
\end{figure*}
%%%%%%%%%%%%%%%%%%%%%%%%%%%%%%%%

First of all, the mixing angle $\theta$ is constrained from the $Z$-decay width measurements \cite{Electroweak:2003ram,Tanabashi:2018oca}, as it induces the invisible decay of $Z$ into $S_1^{{\epsilon}}S_1^{{-\epsilon}}$ whose amplitude  is directly proportional to $\cos 2\theta$. To satisfy this constraint, we choose the mixing angle to be $\theta\simeq \pi/4 $ \cite{Herms:2023cyy}.

Since the mass splitting between $S_1^{{\epsilon}}$ and the other scalar states $\{S_2^{{\epsilon}},\phi_1^{1+{\epsilon}},\phi_2^{1-{\epsilon}}\}$ is required to be large, the new contributions to the EW-precision observables, particularly to the $T$-parameter ~\cite{Peskin:1990zt,Peskin:1991sw} could be significant in this scenario. In the limit  $\epsilon \ll 1$, the corresponding contributions to the $T$-parameter is given by \cite{Herms:2023cyy}:
\begin{align} 
T  &\simeq  \dfrac{1}{8\pi^2 \alpha_{\rm em}(M_Z) v^2 } \bigg\{ c_{\theta}^2  \mathcal{F}(m_{\phi_1^{+}}^2,m_{S_1}^2)
\nonumber\\&
+ s_{\theta}^2  \mathcal{F}(m_{\phi_2^{+}}^2,m_{S_1}^2) + s_{\theta}^2  \mathcal{F}(m_{\phi_1^{+}}^2,m_{S_2}^2)
\nonumber\\&
+c_{\theta}^2  \mathcal{F}(m_{\phi_2^{+}}^2,m_{S_2}^2)
    -s^2_{2\theta}\mathcal{F}(m_{S_1}^2,m_{S_2}^2) \bigg\} \,, \label{eq:T}
\end{align}
where  $\mathcal{F}$ is defined as
\begin{equation} \label{Fdef}
    \mathcal{F}(m_1^2,m_2^2) \  \equiv \  \frac{1}{2}(m_1^2+m_2^2) -\frac{m_1^2m_2^2}{m_1^2-m_2^2}\ln\left(\frac{m_1^2}{m_2^2}\right)\,.
\end{equation}
For a mass hierarchy of the form $m_{S_1}\ll m_{S_2}\simeq m_{\phi_1}\simeq m_{\phi_2}$, one obtains $T\propto \cos^2{2\theta}$, which is naturally suppressed for the chosen value of  $\theta\simeq \pi/4 $.

Now, we investigate the implications of the quartic couplings.
With the choice  of the mass hierarchy $m_{S_1}\ll m_{S_2}\simeq m_{\phi_1}\simeq m_{\phi_2}$ and $\theta\simeq \pi/4 $ , one obtains the following form for  
\begin{align}
\label{eq:lambda'}
\lambda'_{H\phi_{1,2}}\simeq-\lambda_{H\phi_{1}\phi_2}\simeq\frac{m^2_{S_1}-m^2_{S_2}}{v^2}.
\end{align}
On the other hand, the combination of the quartic couplings, $(\lambda_{H \phi_1}+\lambda_{H \phi_2}+\lambda'_{H \phi_1}+\lambda'_{H \phi_2}-2\lambda_{H \phi_1\phi_2})$ is required to be small, as it induces the invisible decay of the SM Higgs into $S_1^{{\epsilon}}S_1^{{-\epsilon}}$, which is constrained by the current bounds on the branching ratio of this decay process
\begin{align}
\label{eq:inv_h}
        \mathrm{Br}_{\rm inv}\leq 
        \begin{cases}
            0.145 & \mathrm{ATLAS}\, \mbox{\cite{ATLAS:2022yvh}}, \\
            0.18 & \mathrm{CMS}\, \mbox{\cite{CMS:2022qva}}.
        \end{cases}
\end{align}
This leads to the following relation
\begin{align}
\label{eq:lambda}
\lambda_{H\phi_{1}}+\lambda_{H\phi_{2}}& \simeq 4\lambda_{H\phi_{1}\phi_2}\simeq \frac{4(m^2_{S_2}-m^2_{S_1})}{v^2}.
\end{align}
However, this condition has non-trivial consequences on the $h\rightarrow \gamma \gamma$ process, as the new physics contributions to this decay mode depend on  $\lambda_{H\phi_{1}}$ and $\lambda_{H\phi_{2}}$ in our model.  To see this, the relevant interactions pertaining to this decay mode are given below
 \begin{align}
    &V\supset \lambda_{H \phi_1} v (h|\phi_1^{1+{\epsilon}}|^2) + \lambda_{H \phi_2} v (h|\phi_2^{1-{\epsilon}}|^2).
\end{align}
Through these interactions, the scalars $\{\phi_1^{1+{\epsilon}},\phi_2^{1-{\epsilon}}\}$ contribute to the $h\rightarrow \gamma \gamma$ process at the one-loop level.
Such new physics contributions alter the Higgs signal strength to $\gamma \gamma$, which is defined as $R_{\gamma\gamma}=\frac{\mathrm{Br}(h\rightarrow \gamma\gamma)}{\mathrm{Br}(h\rightarrow \gamma\gamma)_{\rm SM}}$.
In the limit $\epsilon\ll 1$ and $m_{\phi_1}\simeq m_{\phi_2}\simeq \mathcal{O}(100)\,\GeV$, we obtain $R_{\gamma\gamma}\simeq 0.8$.  This is consistent with  experimental measurements at the 3 sigma level~\cite{ATLAS:2022tnm,CMS:2022wpo}.

In addition to these requirements on the quartic couplings, they also need to satisfy various conditions to ensure the stability of the scalar potential. The corresponding conditions are given by \cite{Hagedorn:2018spx}
\begin{align}
    &\lambda_{H\phi_1}\geq 0, \quad \lambda_{H\phi_1} + \lambda'_{H\phi_1}\geq 0, \label{eq:cond1} \\
    &\lambda_{H\phi_2}\geq 0, \quad \lambda_{H\phi_2} + \lambda'_{H\phi_2}\geq 0, \label{eq:cond2}\\
    & \sqrt{(\lambda_{H\phi_1} + \lambda'_{H\phi_1})(\lambda_{H\phi_2} + \lambda'_{H\phi_2})}-\lambda_{H\phi_1 \phi_2}\geq 0.  \label{eq:cond3}
\end{align}
We checked whether the above conditions are consistent with Eqs.~\eqref{eq:lambda'} and \eqref{eq:lambda} and found that they are compatible.

%%%%%%%%%%%%%%%%%%%%%%%%%%%%%%%%%%
\begin{figure*}[t!]
$$
\includegraphics[width=0.425\textwidth]{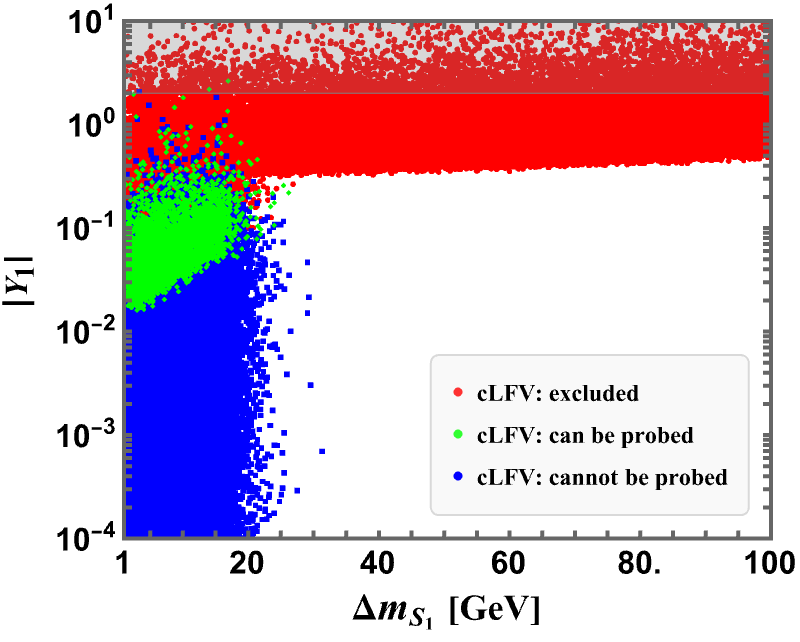}
\hspace{0.45in}
\includegraphics[width=0.45\textwidth]{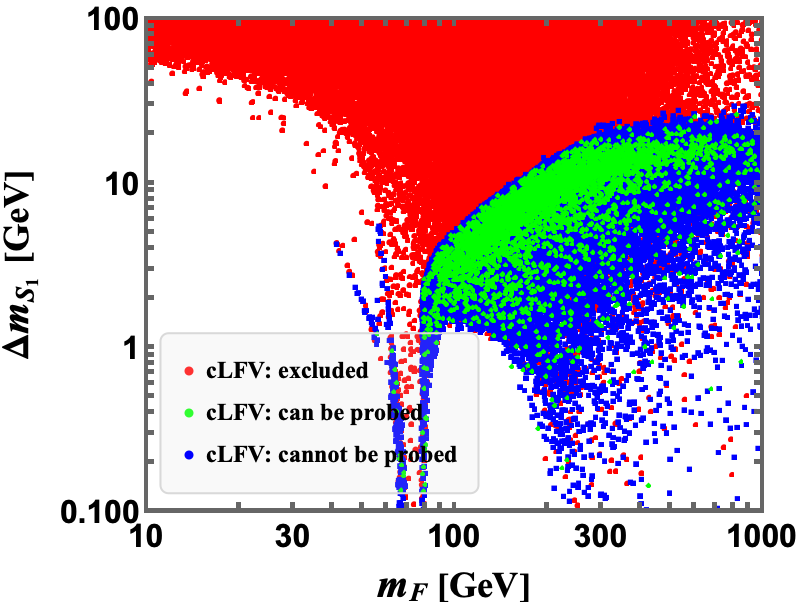}
$$
\caption{Left (right) panel: parameter space in $\Delta m_{S_1}\equiv m_{S_1}-m_F$ vs. $|Y_1|$ $(m_F$ vs. $\Delta m_{S_1})$ plane consistent with the Planck relic density constraint. Here, we set $S_1^{\epsilon}$ to be the coannihilation partner of DM.
The horizontal gray-shaded band denotes the region excluded by the perturbativity bound on the Yukawa coupling $|Y_1|<2$.
The other color codes represent the same as in Fig.~\ref{fig:lightDM}.
}\label{fig:heavyDM_S1}
\end{figure*}
%%%%%%%%%%%%%%%%%%%%%%%%%%%%%%%%%%%

The scalars $\{\phi_1^{1+{\epsilon}},\phi_2^{1-{\epsilon}}\}$ contribute to charged lepton flavor violation (cLFV) processes via the Yukawa couplings $Y_{1,2}$. In the limit  $\epsilon \ll 1$, 
the corresponding contributions to $\ell_\alpha\to \ell_\beta\gamma $ can be written as \cite{Herms:2023cyy,Hagedorn:2018spx}
\begin{align}
\label{eq:lfvBRformula}
&\frac{\mathrm{BR}\left(\ell_\alpha\to \ell_\beta\gamma\right)}{\mathrm{BR}\left(\ell_\alpha\to \ell_\beta\nu_\alpha\overline\nu_\beta\right)}
\simeq \frac{3\alpha_\mathrm{EM}}{64\pi G^2_F}
\times 
\nonumber\\ &~~~~~~~~~~~~~~
\bigg\{
\left|\frac{Y_{1,\beta}^*Y_{1,\alpha}}{m^2_{\phi_1}}f\left( \frac{m_{F}^2}{m^2_{\phi_1}}\right) + \frac{Y_{2,\beta}^*Y_{2,\alpha}}{m^2_{\phi_2}}f\left( \frac{m_{F}^2}{m^2_{\phi_2}}\right)\right|^2   
\bigg\},
\end{align}
with the auxiliary function
\begin{align}
f(x)=\frac{1-6x+3x^2+2x^3-6x^2\ln x}{6(1-x)^4} \,.
\end{align}
The current bounds (and future sensitivities) on the relevant branching ratios at $\SI{90}{\percent}$ CL are given by
\begin{align}
\label{eq:cLFV}
&\mathrm{BR}(\mu\to e\gamma) < 3.1\times 10^{-13}  \,\, (6.0\times 10^{-14}) \, \, \, \mbox{\cite{MEGII:2023ltw}}, \nonumber \\
&\mathrm{BR}(\tau\to e\gamma) < 3.3\times 10^{-8}\, \mbox{\cite{BaBar:2009hkt}} \,\, (3\times 10^{-9}\, \mbox{\cite{Belle-II:2018jsg}}), \nonumber \\
&\mathrm{BR}(\tau\to \mu\gamma) < 4.2\times 10^{-8}\, \mbox{\cite{Belle:2021ysv}} \, \, (10^{-9}\, \mbox{\cite{Belle-II:2018jsg}}), 
\end{align}
and impose stringent limits on the Yukawa couplings.

Taking into account all these constraints,
we explore the parameter space that can explain the correct relic abundance for DM. For this, we perform a random scan over the ranges $m_F<m_{S_1}\in (10 \, \MeV - 10 \,\GeV)$ and $m_{S_2}=m_{\phi_1}=m_{\phi_2} \in(110 \,\GeV- 500\,\GeV)$. The Yukawa couplings are fixed through the parametrization given in Eq.~\eqref{eq:YukpNH} for NH and Eq.~\eqref{eq:YukpIH} for IH. The entries of the PMNS matrix are determined by varying over the $3\sigma$ range of neutrino observables taken from NuFIT 5.3 \cite{Esteban:2020cvm} under the assumption of flat priors. We also fix the hierarchy of these couplings to be $|Y_1|>|Y_2|$. 
Since one requires sizeable Yukawa couplings to generate a sufficiently large contribution to $\sigma v$ \cite{Herms:2023cyy}, we vary $k$ in the range $(10^2,10^8)$. From this data sample, we select the data points that satisfy the DM relic density constraint \cite{Planck:2018vyg}. The results of this analysis are shown in the DM mass ($m_F$) versus Yukawa coupling ($|Y_1|$) plane for both NH (left panel) and IH (right panel) in Fig.~\ref{fig:lightDM}. In this figure, the red-colored data points are excluded by the cLFV constraints, whereas the green (blue)- colored data points are allowed and can (not) be probed in future cLFV experiments.
Notice that, for larger DM masses, sizeable values of Yukawa couplings are required to be consistent with the DM relic density constraint. However, such large values of Yukawa couplings are excluded by the cLFV constraints for $m_F\gtrsim 0.8 \, \GeV$. Moreover, it is also worth mentioning that these constraints are stronger for the IH of neutrino masses as already DM masses $m_F\gtrsim 0.5 \, \GeV$ are ruled out.

%%%%%%%%%%%%%%%%%%%%%%%%%%%%%%%%%%%%%%%%%%%%
\begin{figure*}[t!]
$$
\includegraphics[width=0.425\textwidth]{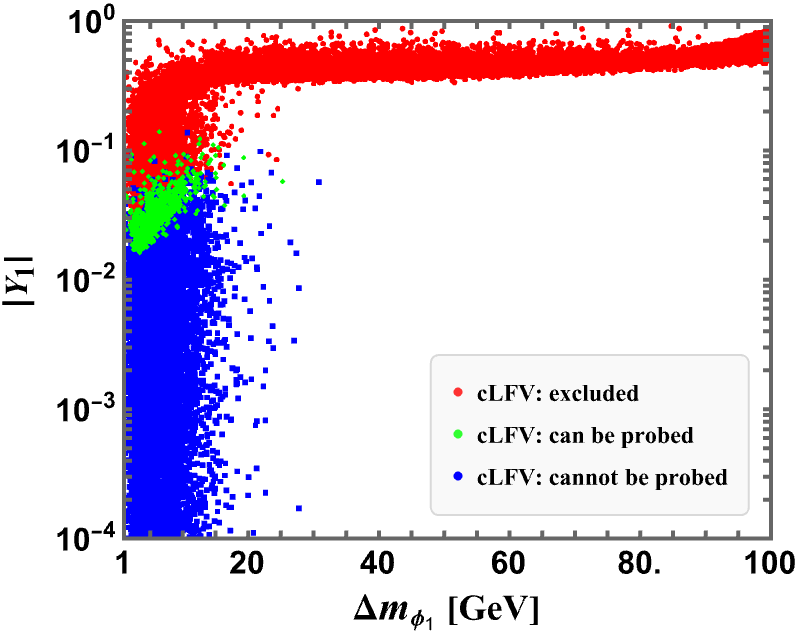}
\hspace{0.45in}
\includegraphics[width=0.45\textwidth]{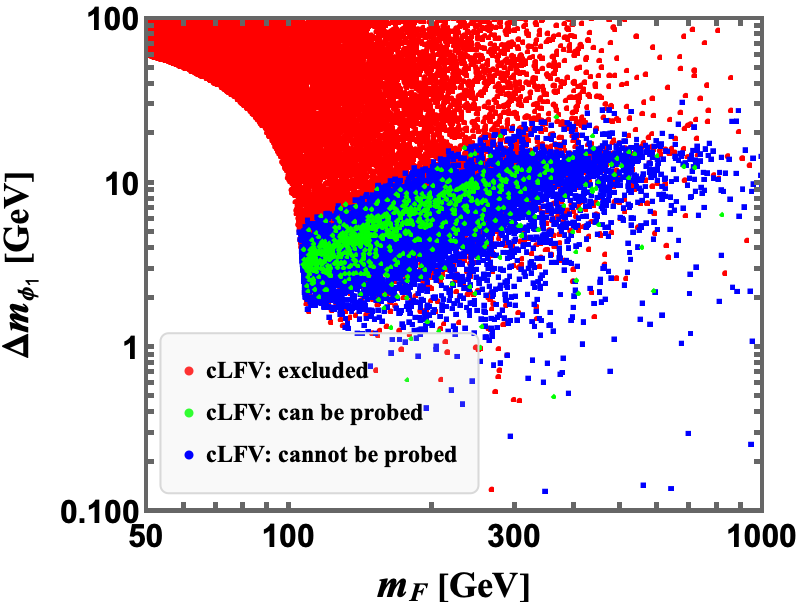}
$$
\caption{Left (right) panel: parameter space in $\Delta m_{\phi_1}\equiv m_{\phi_1}-m_F$ vs. $|Y_1|$ $(m_F$ vs. $\Delta m_{\phi_1})$ plane consistent with the Planck relic density constraint. 
Here, we set $\phi_1^{1+\epsilon}$ to be the coannihilation partner of DM. The color codes represent the same as in Fig.~\ref{fig:lightDM}.
}\label{fig:heavyDM_Phi}
\end{figure*}
%%%%%%%%%%%%%%%%%%%%%%%%%%%%%%%%%%%%%%%%%

\textbf{\emph{Heavy dark matter}.--}
As discussed above, for larger DM masses, its annihilation into SM leptons via the $t$-channel processes is  excluded through the cLFV constraints. 
However, the coannihilations with the new scalars are less severely constrained by these constraints, as they only depend on the square of the Yukawa couplings, while coannihilations of the new scalars into, e.g., gauge bosons are independent of the Yukawas.
Hence, for the heavy DM scenario, the coannihilation processes can become important for explaining the correct relic abundance. 
To generate a sufficiently large contribution to $\sigma v$ through these processes, the mass splitting between the DM and the coannihilation partner is required to be small \cite{Griest:1990kh}. In our framework, such small mass splittings are allowed.
For the numerical study, we consider two scenarios. In the first scenario, the \textit{milli-charged} scalar $S_{1}^{\epsilon}$ is considered to be the next-to-lightest new particle, whereas in the second, $\phi_1^{1+\epsilon}$ is considered to be the next-to-lightest particle. For the analysis, we use the following toolchain. {\tt{minimal-lagrangians}}~\cite{May:2020sod} is used to create the model input for {\tt{SARAH 4.15.1}}~\cite{Staub:2013tta} which in turn generates the  {\tt{SPheno 4.0.5}}~\cite{Porod:2011nf} module as well as the {\tt{CalcHEP}}~\cite{Belyaev:2012qa} model used to calculate the particle spectrum and the cLFV observables (Eq. \eqref{eq:cLFV}), the $S$, $T$ and $U$ parameters and the files for {\tt{micrOMEGAs 5.3.41}}~\cite{Belanger:2018ccd} used to calculate the relic density.

In Fig.~\ref{fig:heavyDM_S1} (left panel), we show the parameter space for the first scenario consistent with the Planck relic density constraint \cite{Planck:2018vyg} in the $\Delta m_{S_1}\equiv m_{S_1}-m_F$ versus $|Y_1|$ plane using the same color codes  as in Fig.~\ref{fig:lightDM} for the NH of neutrino masses. The associated scan is over the ranges
\begin{align}
    &m_F \in [\SI{10}{\giga\electronvolt},\SI{e5}{\giga\electronvolt}]\, , \nonumber \\
    &m_{S_1}-m_F < \SI{100}{\giga\electronvolt} \, \mathrm{with} \, m_{S_2} \in  [m_{S_1},\SI{e5}{\giga\electronvolt}] \, , \nonumber \\
    &m_{\phi_1},m_{\phi_2}  \in [\max(m_{S_1},\SI{110}{\giga\electronvolt}),\SI{e5}{\giga\electronvolt}] \, ,\nonumber \\ 
    &\lambda_{H\phi_i} \in [\max(-\lambda'_{H\phi_i} ,10^{-10}),\sqrt{4 \pi}] \, ,\nonumber \\
     &\lambda_{H \phi_1 \phi_2}  \in [10^{-8},\sqrt{4 \pi}] \,\mathrm{and}\, k\in [1,10^{10}]
     \label{eq:ranges}
\end{align}
under the assumption of logarithmic priors and
using the three-flavor neutrino oscillation parameters taken from the global fit of the \texttt{NuFIT 5.2} collaboration \cite{Esteban:2020cvm}. The vacuum stability conditions in Eqs.~\eqref{eq:cond1} to \eqref{eq:cond3} are enforced as well. 
Notice that, for the larger mass splittings, larger values of the Yukawa couplings are required to generate the correct relic abundance, which are severely constrained by the cLFV constraints. On the other hand, for lower mass splittings, the Yukawa couplings are allowed to be much smaller since the coannihilation processes become the dominant annihilation mode in this regime. In the figure shown in the right panel, we show the parameter space for the same scenario in the DM mass versus $\Delta m_{S_1}$ plane.  Notice that the region where $m_F \gtrsim 40\,\GeV$ is consistent with all the constraints including the cLFV bounds.  However, DM masses below $40 \,\GeV$ are not favored from these constraints, since both coannihilation and annihilation of $S_1^{\epsilon}$ are either suppressed or kinematically forbidden in this region.

In Fig.~\ref{fig:heavyDM_Phi}, the parameter space for the second scenario involving dominantly $\phi_1^{1+\epsilon}$ coannihilations is shown. The corresponding scan ranges are identical to those in Eq.~\eqref{eq:ranges} except with the roles of $S_1^{\epsilon}$ and $\phi_1^{1+\epsilon}$ being interchanged, i.e.,
\begin{align}
     &m_{\phi_1}-\max(110,m_F) < \SI{110}{\giga\electronvolt} \nonumber \\
     &m_{S_1} \in [m_{\phi_1},\SI{e5}{\giga\electronvolt}]
\end{align}
As one can see from this figure, DM masses below 100 GeV are excluded by the cLFV constraints since the coannihilation processes are suppressed in this region because of $m_{\phi_1}\gtrsim 110 \,\GeV$. 

%%%%%%%%%%%%%%%%%%%%%%%%%%%%%%%%%%%%%%%%%%%%%
\section{Implications on Neutrinoless double beta decay \label{sec:nubeta}}
%%%%%%%%%%%%%%%%%%%%%%%%%%%%%%%%%%%%%%%%%%%%%
Our framework predicts neutrinoless double beta decay at an observable level. To demonstrate this, first, we calculate the contribution to the effective neutrino mass $m_{\beta\beta}$ in our model. Since the lightest neutrino is massless in our setup,    $m_{\beta\beta}$ takes the following form \cite{Herms:2023cyy}:
\begin{align}
    m_{\beta\beta}^{NH} &=\left|m_2 s_{12}^2 c_{13}^2 + m_3 s_{13}^2 e^{i\alpha}  \right|, \nonumber \\
    m_{\beta\beta}^{IH}  &=\left|m_1 c_{12}^2 c_{13}^2 + m_2 s_{12}^2 c_{13}^2 e^{i\alpha}  \right|,
    \label{eq:onbb}
\end{align}
where $s_{ij}\equiv\sin{\theta_{ij}}$ ($c_{ij}\equiv\cos{\theta_{ij}}$), and $\theta_{ij}$ and $\alpha$ denote the neutrino mixing angles and Majorana phase, respectively. As a consequence of neutrino masses being generated radiatively in our framework, $m_{\beta\beta}$ given in Eq. \eqref{eq:onbb} will be modified by the loop effect \cite{Rodejohann:2019quz}. In the limit where the scale of neutrino self-energy (denoted as $\Lambda$) is greater than the momentum transfer of the $0\nu\beta\beta$ process (denoted as $p$), the modification of $m_{\beta\beta}$ is given by \cite{Rodejohann:2019quz,Herms:2023cyy}
\begin{equation}
    \label{eq:onbbL}
     m_{\beta\beta}  \longrightarrow m_{\beta\beta} \left(1+\frac{p^2}{\Lambda^2}\right),
\end{equation}
where $\Lambda$ is given by \cite{Herms:2023cyy}
\begin{align}
&\Lambda^2=2 \frac{\mathcal{G}[m^2_{S_2}]-\mathcal{G}[m^2_{S_1}]}{\mathcal{H}[m^2_{S_2}]-\mathcal{H}[m^2_{S_1}]}.
%\\
\end{align}
Here, the loop functions are defined as
\begin{align}
&\mathcal{G}[m^2_{S}]\equiv\frac{m_{S_i}^2\ln\left( \frac{m_{S}^2}{m^2_F} \right) }{m_{S}^2-m_F^2},
\\
&\mathcal{H}[m^2_{S}]\equiv\frac{m^4_{F}-m^4_{S}+2m^2_{S}m^2_{F}\ln{\dfrac{m^2_{S}}{m^2_{F}}}}{(m^2_{S}-m^2_{F})^3}.
\end{align}
Notice from Eq.~\eqref{eq:onbbL} that the loop effect becomes significant when $\Lambda \simeq \mathcal{O}(p)$. This implies that the loop effect is more significant for the light DM scenario compared to the heavy DM case since the typical scale of momentum transfer of $0\nu\beta\beta$ process is $\mathcal{O}(100)\,\MeV$ \cite{Simkovic:2018hiq, Shimizu:2017qcy, Graf:2023dzf}.

\begin{figure}[thb!]
$$
\includegraphics[width=0.4\textwidth]{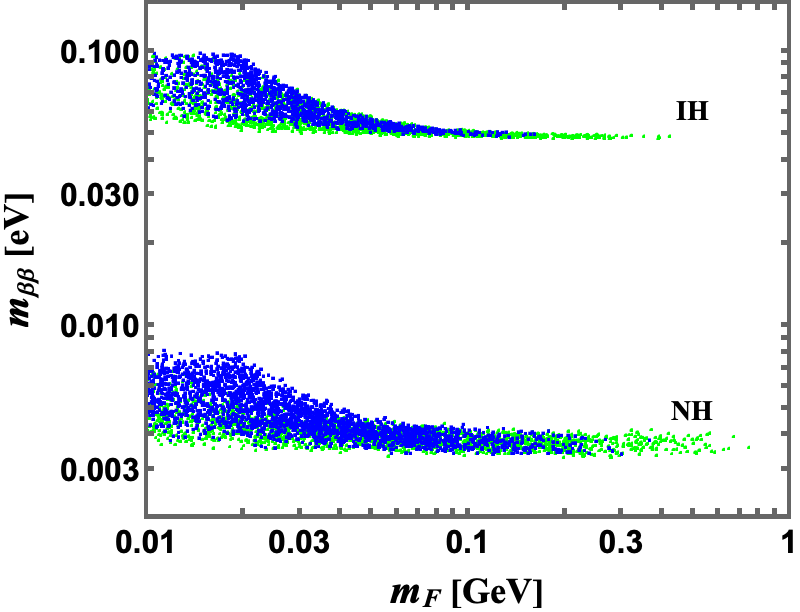}
$$
\caption{Parameter space in DM mass ($m_F$) vs. effective neutrino mass ($m_{\beta\beta}$) for both NH and IH of neutrino masses.  For concreteness, we fix $p=200 \MeV$ and $\alpha=0$.
The blue and green color codes denote the same as in Fig.~\ref{fig:lightDM}.
}\label{fig:onbb}
\end{figure}

In Fig.~\ref{fig:onbb}, we show the parameter space in DM mass versus the effective neutrino mass plane consistent with the Planck relic density constraint and the cLFV constraints. For concreteness, we fix $\alpha=0$ and $p=200\,\MeV$. Notice that from this figure, the contribution to $m_{\beta\beta}$ is enhanced for the lower DM masses, whereas for the higher masses, the effect is negligible.

%%%%%%%%%%%%%%%%%%%%%%%%%%%%%%%%%%%%%%%%%%%%%
%%%%%%%%%%%%%%%%%%%%%%%%%%%%%%%%%%%%%%%%%%%%%

%%%%%%%%%%%%%%%%%%%%%%%%%%%%%%%%%%%%%%%%%%%%%%%%%%%%%%
\medskip
%\textbf{\emph{Conclusions}.--}
\section{Conclusions \label{sec:conclusion}}
In this work, we have presented a  radiative neutrino mass mechanism in which the particles within the loops carry small electric charges. Contrary to the conventional \textit{scotogenic setup},  this scheme does not require the ad-hoc imposition of any new symmetry for stabilizing the DM candidate since its stability is ensured through its minuscule electric charge. The minimal UV completion of this scenario requires one generation of a \textit{milli-charged} fermion $F$ and two EW-scalar doublets $\phi_1$ and $\phi_2$ in addition to the SM particles.  

We have investigated the DM phenomenology associated with the lightest \textit{milli-charged} particle $F$, for both freeze-in and freeze-out scenarios finding that the freeze-in mechanism is not compatible with the neutrino mass mechanism in our model while  the freeze-out mechanism works successfully. We explored the thermal relic scenario from sub-GeV to TeV DM masses. In the sub-GeV mass regime, the DM candidate mainly self-annihilates into neutrinos via the leptonic portal whereas, for the heavy DM scenario,  coannihilation processes set the relic abundance.    

We have also discussed various other phenomenological implications of our model, such as lepton flavor-violating observables, electroweak precision observables, and neutrinoless double beta decay. We found that a vast region of parameter space can be probed in future cLFV experiments. Additionally, our model also predicts neutrinoless double beta decay at an observable level.

\begin{acknowledgments}
{\textbf {\textit {Acknowledgments.--}}} We would like to thank A. Pukhov for correspondence on the implementation of \textit{milli-charged} DM into {\tt{micrOMEGAs}}. This work has been supported by the DFG through the Research Training Network 2149 “Strong and Weak Interactions - from Hadrons to Dark Matter”.
\end{acknowledgments}
\bibliographystyle{utphys}
\bibliography{reference}
\end{document}